\documentclass{article}
\usepackage[margin=1in]{geometry}    
\usepackage{graphicx} 
\usepackage{graphicx}
\usepackage{dcolumn}
\usepackage{bm}
\usepackage[utf8]{inputenc}
\usepackage{graphicx}
\usepackage{amssymb}
\usepackage{array}
\usepackage{wrapfig}
\usepackage{tabularx}
\usepackage{hyperref}
\usepackage[
  backend=biber,
  style=numeric-comp,
  sorting=none
]{biblatex}
\usepackage{float}
\usepackage{breqn}
\usepackage{tikz}
\usetikzlibrary{automata}
\expandafter\let\csname equation*\endcsname\relax
\expandafter\let\csname endequation*\endcsname\relax
\usepackage{amsmath}
\usepackage{multirow}
\usepackage[utf8]{inputenc}
\usepackage{dirtytalk}
\usepackage{tabularx}
\usepackage{subcaption}
\usepackage{tikz}
\usetikzlibrary{positioning, shapes.multipart, arrows.meta, fit}
\title{Relational Time as a Stochastic Variable in ADM Gravity}
\author{Pradosh keshav MV \\
Department of Physics and Electronics, \\
Christ University, Bangalore, India, 560029}

\begin{document}

\maketitle
\begin{abstract}
The absence of a fundamental time parameter in canonical quantum gravity motivates the search for internal clocks that can define evolution relationally. While classical deparametrization schemes provide formal solutions, they often rely on semiclassical limits or fixed foliations that break general covariance. In this work, we construct a canonical framework where time emerges dynamically as a stochastic degree of freedom, identified with a massless scalar field whose conserved momentum current defines a relational foliation. Crucially, we show that unresolved transverse-traceless graviton modes induce stochastic corrections to the Hamiltonian constraint, leading to a Langevin-type extension of the Wheeler–DeWitt equation. The resulting dynamics preserve the hypersurface deformation algebra in an ensemble sense and give rise to probabilistic evolution within a fully covariant setting. This is of considerable interest towards a minimal mechanism for embedding quantum fluctuations into the constraint structure of gravity, yielding a non-semiclassical and internally consistent realization of temporal decoherence, and emergent classicality in quantum cosmology.
\end{abstract}

\section{introduction}
 
In canonical quantum gravity, the problem of time presents a structural and conceptual challenge. The Wheeler–DeWitt (WD) equation, arising from the quantization of the Hamiltonian constraint, eliminates any explicit temporal parameter, reducing evolution to internal correlations among observables. While consistent with diffeomorphism invariance, this timeless formalism leaves open the question of how temporal experience — and quantum evolution itself — arises from within the theory. This tension was already implicit in the earliest attempts to formulate a dynamical notion of time in general relativity. The Baierlein–Sharp–Wheeler (BSW) action \cite{Baierlein1962} demonstrated that time could, in principle, be eliminated in favor of intrinsic spatial geometry, yielding a Jacobi-type square-root action where evolution is encoded in the 3-metric itself. This geometrical approach was retained in the canonical formalism developed by DeWitt and Misner \cite{DeWitt1967, Misner1969}, but the resulting WD equation nonetheless preserved a frozen, time-independent structure. Page and Wootters’ framework \cite{page1983evolution} conditions probabilities on quantum clock readings, but neglects feedback from clock fluctuations — a point emphasized by Gambini, Porto, and Pullin \cite{gambini2007fundamental} in their analysis of relational decoherence.

Various proposals have attempted to resolve this, such as relational time models \cite{kuchavr1973canonical, rovelli1996relational, page1983evolution}, deparametrization through scalar fields \cite{brown1995dust}, and semiclassical approximations while promoting a dynamical field to serve as a clock. Scalar fields are especially promising in this regard, serving simultaneously as dynamical matter and relational timekeepers \cite{Gielen2021}. However, when formulated in the Arnowitt–Deser–Misner (ADM) framework, the quantization of the scalar field becomes entangled with gravitational degrees of freedom, obstructing its interpretation as an independent temporal parameter \cite{Landsman1995}. The evolution they define typically depends on specific gauge choices, which introduce foliation dependence and undermine full diffeomorphism invariance. This issue is evident in modified gravity scenarios, such as Horava–Lifshitz and cuscuton models, where explicit foliation leads to pathological effects including ghost condensation and “eternity skins” near horizons \cite{magueijo2024black, blas2011models}. Even in relational frameworks, quantum clocks defined by scalar fields often suffer from ambiguity in their conditional probabilities due to backreaction and field fluctuations \cite{anderson2012problem}. Although semiclassical approaches have been used to stabilize clock dynamics, these typically treat gravitational backreaction only at the level of expectation values, ignoring the fluctuations essential for a consistent quantum theory of spacetime.

Recent developments in stochastic gravity have reopened the possibility that time itself may acquire a probabilistic structure. Elze and Schipper \cite{elze2002time} were among the first to propose stochastic time variables in a relational context, although their model lacked constraint preservation and diffeomorphism invariance. In a different setting, Hu and Verdaguer developed a stochastic semiclassical gravity framework in which noise emerges from quantum-field fluctuations, inducing stochastic corrections to the Einstein equations \cite{hu2008stochastic, hu2003stochastic}. While the Hu–Calzetta \cite{calzetta1997stochastic, calzetta1988nonequilibrium, calzetta1994noise, calzetta2001coarse, calzetta2009nonequilibrium}  framework emphasizes a foundational derivation based on the influence functional and open quantum systems, the Hu–Verdaguer approach refines this into a computationally tractable formalism applicable to cosmology and black hole spacetimes via the Einstein–Langevin equation. However, both approaches presuppose a classical background geometry and do not capture fully quantum backreaction effects. More recently, Erlich’s composite gravity program \cite{erlich2018stochastic, erlich2022first} introduced stochastic fluctuations in matter fields to construct a dynamically emergent, probabilistic spacetime geometry. Although these models offer intriguing insights into how stochasticity might underlie quantumness and gravity, they often leave unresolved how noise interacts with the constraint structure of canonical gravity. Cable’s recent second-order stochastic formalism \cite{Cable2022Second-order} presents a step forward showing that quantum fluctuations can generate controlled foliation variability while preserving relational observables. Yet it remains unclear whether stochasticity can be consistently implemented in the scalar clock sector in a way that preserves the closure of the Hamiltonian constraints and retains covariance at the quantum level.

In this work, we develop a stochastic-clock framework in which the scalar field is treated not as a deterministic internal time, but as a coarse-grained, stochastic degree of freedom whose fluctuations encode unresolved gravitational dynamics. These fluctuations are not introduced ad hoc; rather, they arise from a physical coarse-graining of the transverse–traceless (TT) graviton modes within a canonical path integral \cite{pawlowski2021quantum}. This is analogous to integrating out short-wavelength gravitational degrees of freedom in Wilsonian effective field theory. By imposing a divergence-free condition on the scalar field momentum, we define a dynamically conserved slicing that replaces externally fixed foliations with an internally generated, relational time variable. This leads to a stochastic deformation of the Hamiltonian constraint which modifies the WD equation while preserving the closure of the hypersurface deformation algebra in an ensemble-averaged sense. Building on Isham’s call for variational reparametrization invariance \cite{Isham1993}, our construction allows stochastic fluctuations in the clock sector without violating the canonical constraint structure. While we do not claim to resolve the unitarity problem outright, we demonstrate that relational observables defined with respect to a stochastic clock can evolve probabilistically in a way that remains consistent with the underlying symmetries of canonical quantum gravity. Thus, time is no longer introduced as a classical parameter or emergent gauge label, but appears as a physical quantum field with intrinsic stochasticity — embedded within the dynamics of spacetime itself.

The structure of the paper is as follows. In Sec. II, we review the role of minimally coupled scalar fields as internal clocks in canonical general relativity. We argue that the presence of unresolved (UV) gravitational degrees of freedom—specifically, coarse-grained transverse-traceless graviton modes— introduces fluctuations in the clock momentum that break exact divergence freedom. In Sec. III, we construct the corresponding stochastic canonical framework by incorporating noise terms into the diffeomorphism and Hamiltonian constraints. This leads to a deformation of the WD equation in the form of a Langevin-type evolution, where the stochastic clock drives relational dynamics. Sec. IV presents an explicit realization of this formalism in a homogeneous and isotropic minisuperspace cosmology. In this setting, the stochastic deformation modifies the evolution of the wavefunction through a diffusion term, affecting semiclassical trajectories and relational observables. In Sec. V, we discuss the implications of this framework with its scope and future extensions.

\section{Relational Time and the Survival of Clocks in Canonical Gravity}
In this section, we begin by reviewing how relational clocks emerge within the ADM formalism, and how scalar fields can be used to define physically meaningful slicings of spacetime. We then analyze the gauge fixing associated with such clocks, derive the resulting constraint structure, and show how this leads to a Schrödinger-type evolution in terms of a scalar clock.

\subsection{Relational Clocks and the ADM Foliation}

In canonical general relativity, the absence of a preferred temporal parameter is inherent to the theory's general covariance. The ADM formalism addresses this by decomposing spacetime into a foliation of spatial hypersurfaces \( \{ \Sigma_t \} \), labeled by an arbitrary coordinate parameter \( t \). The four-dimensional spacetime metric is expressed as:
\begin{equation}
    ds^2 = N^2 dt^2 - \gamma_{ij}(dx^i + N^i dt)(dx^j + N^j dt),
\end{equation}
where \( N \) is the lapse function, \( N^i \) is the shift vector, and \( \gamma_{ij} \) is the spatial 3-metric induced on each slice \( \Sigma_t \). The lapse and shift encapsulate the arbitrariness of foliation, embodying the freedom inherent in four-dimensional diffeomorphism invariance.

However, the coordinate time \( t \) lacks intrinsic physical meaning. To recover a notion of physical temporal evolution, one introduces a dynamical internal variable that parametrizes progression: a relational clock. A widely adopted choice is a massless scalar field \( \phi(x) \), whose level sets \( \phi(x) = \tau \) define physical hypersurfaces. Provided that the gradient \( \partial_\mu \phi \) is timelike, these level sets define spacelike hypersurfaces with a normal vector:
\begin{equation}
    n_\mu = \frac{\partial_\mu \phi}{\sqrt{g^{\alpha\beta} \partial_\alpha \phi \partial_\beta \phi}}.
\end{equation}

Within this framework, the scalar field \( \phi(x) \) does not merely represent a matter degree of freedom, but establishes a slicing of spacetime. This procedure replaces arbitrary coordinate time with evolution relative to a physical, dynamical field \( \phi \). Relational slicing based on scalar fields has been successfully implemented in various quantum gravity approaches, including group-field theory and deparametrized models of loop quantum cosmology.

Nevertheless, relational clocks are predicated upon the assumption that \( \phi \) evolves smoothly and monotonically, unaffected by its coupling to the gravitational field. This idealization holds in semiclassical regimes but becomes questionable in a fully quantum gravitational context, where the clock field interacts with unresolved gravitational degrees of freedom.

This observation motivates extending the classical notion of relational clocks to accommodate such interactions. Specifically, we propose that coarse-graining over unresolved gravitational modes induces stochastic fluctuations in the clock field's evolution. Consequently, the relational time defined by \( \phi \) acquires intrinsic probabilistic characteristics, leading to what we term a \emph{stochastic clock}.

In the following sections, we develop this concept systematically. By analyzing the constraint structure and gauge-fixing conditions associated with scalar clocks, we reveal how the lapse function becomes dynamically tied to the clock momentum. We then formalize the stochastic clock framework, treating time itself as a stochastic process sourced by coarse-grained gravitational fluctuations.

\subsection{Relational Clocks and the ADM Foliation}

In canonical general relativity, the dynamical content of the gravitational field is most naturally described via a foliation of spacetime into a family of spatial hypersurfaces. Let $(M, g_{\mu\nu})$ be a globally hyperbolic four-dimensional Lorentzian manifold, equipped with a foliation $\{\Sigma_t\}_{t \in \mathbb{R}}$ by spacelike hypersurfaces diffeomorphic to a fixed 3-manifold $\Sigma$. The foliation can be specified by an embedding map
\begin{equation}
\iota : \Sigma \times \mathbb{R} \rightarrow M,\qquad (x^i, t) \mapsto y^\mu = \iota^\mu(x^i, t),
\end{equation}
which assigns to each spatial point $x^i$ on $\Sigma$ a spacetime point $y^\mu \in M$ lying on the leaf $\Sigma_t$. The evolution vector field along the foliation is given by $Z^\mu = \partial_t \iota^\mu(x, t)$, and can be decomposed into its components normal and tangential to the hypersurfaces as
\begin{equation}
Z^\mu = N n^\mu + N^i e^\mu_i,
\end{equation}
where $n^\mu$ is the future-directed unit normal to $\Sigma_t$, $e^\mu_i = \partial_i \iota^\mu$ are tangent vectors spanning $\Sigma_t$, $N$ is the lapse function, and $N^i$ is the shift vector. This decomposition allows the spacetime metric $g_{\mu\nu}$ to be written in ADM form:
\begin{equation}
ds^2 = g_{\mu\nu} dy^\mu dy^\nu = N^2 dt^2 - \gamma_{ij}(dx^i + N^i dt)(dx^j + N^j dt),
\end{equation}
where $\gamma_{ij}$ is the induced spatial 3-metric on each slice $\Sigma_t$. 

While this construction assumes the existence of a preferred foliation parameterized by a coordinate time $t$, such a coordinate is physically meaningless in a generally covariant theory (see Gielen and Menéndez-Pidal \cite{GielenMenendezPidal2021}). In particular, the arbitrariness of the lapse and shift functions reflects the full four-dimensional diffeomorphism invariance of general relativity. Hence, to recover a notion of temporal evolution, one must introduce a dynamical, internal criterion for slicing spacetime—namely, an observable field whose variation can be interpreted as temporal progression.

Relational time \cite{rovelli1996relational, Rovelli1991, Rovelli1995, RovelliSmolin1994, RovelliAshtekar1991, kuchavr1973canonical, Kuchar1971, Kuchar1981} is one such approach, in which a physical field—typically a scalar field $\phi(x)$—is used to define hypersurfaces via its level sets \cite{nakonieczna2015scalar, nakonieczna2016scalar}. That is, one postulates that the hypersurfaces $\Sigma_\tau$ are defined by $\phi(x) = \tau$, so that $\phi$ plays the role of a relational clock. In this setting, the scalar field $\phi(x)$ is not merely an additional matter degree of freedom, but serves to define a slicing of the spacetime manifold. For instance, \cite{calcinari2025relational} has shown that Page–Wootters–style relational time can be embedded in background‑independent models through explicit matter clock construction. Its gradient $\partial_\mu \phi$ is assumed to be timelike and future-directed such that its level sets $\phi(x) = \tau$ define spacelike hypersurfaces. The unit normal vector to these slices is given by:
\begin{equation}
n_\mu = \frac{\partial_\mu \phi}{\sqrt{g^{\alpha\beta} \partial_\alpha \phi \partial_\beta \phi}}.
\end{equation}
where the evolution vector field $Z^\mu$ becomes aligned with $\partial^\mu \phi$ up to normalization and shift contributions. Such relational slicing has also been successfully implemented within group-field-theory models \cite{wilson2019relational}, where a massless scalar defines “equal relational time” commutation relations. Thus, the scalar field $\phi$ becomes a viable parameter to describe the physical evolution of the system in the Hamiltonian formalism, particularly when no external time coordinate is available.

Notably, defining the foliation dynamically through the field $\phi(x)$ could break the full four-dimensional diffeomorphism symmetry down to a subgroup $\mathrm{Diff}_\phi(M)$ that preserves the slicing \cite{park2015hypersurface}. Infinitesimal diffeomorphisms generated by vector fields $\xi^\mu$ must satisfy the constraint $\mathcal{L}_\xi \phi = \xi^\mu \partial_\mu \phi = 0$ in order to preserve the foliation structure. This residual symmetry acts only tangentially within each hypersurface, i.e., as spatial diffeomorphisms $\xi^i$ such that $\xi^0 = 0$, defining a reduced gauge group $\mathrm{Diff}(\Sigma) \subset \mathrm{Diff}(M)$. Geometrically, the reduction allows one to interpret $(M, \phi)$ as a principal fiber bundle over the space of relational instants, with fibers corresponding to spatial slices $\Sigma_\phi$. This sets the stage for interpreting dynamics in canonical gravity as evolution with respect to the relational time variable $\phi$. It also motivates a gauge-fixing procedure wherein the condition $\phi(x) = \tau$ is treated as a constraint within the Dirac–Bergmann algorithm \cite{dirac1950generalized, anderson1951constraints}.

In the following sections, we will make the gauge-fixing procedure precise by analyzing the consistency of the preferred clock choice and its applications for the lapse function and constraint algebra.

\subsection{Lapse--Momentum Coupling and Schrödinger Evolution in Scalar Time}

Let us begin by recalling that the total Hamiltonian of general relativity minimally coupled to a scalar field is given by
\begin{equation}
H_{\text{tot}} = \int_\Sigma d^3x \left[ N(x) \mathcal{H}(x) + N^i(x) \mathcal{H}_i(x) \right],
\end{equation}
where $\mathcal{H}(x)$ is the Hamiltonian (or scalar) constraint and $\mathcal{H}_i(x)$ is the momentum (or diffeomorphism) constraint. The canonical phase space variables are $(\gamma_{ij}, \pi^{ij}; \phi, \pi_\phi)$, with Poisson brackets
\begin{equation}
\{ \gamma_{ij}(x), \pi^{kl}(y) \} = \delta_{(i}^k \delta_{j)}^l \delta^3(x - y), \qquad
\{ \phi(x), \pi_\phi(y) \} = \delta^3(x - y).
\end{equation}
By imposing the gauge-fixing condition:
\begin{equation}
\chi(x) := \phi(x) - \tau \approx 0, \label{eq:guagecond}
\end{equation}
which identifies the value of the scalar field with a constant label $\tau$ on each hypersurface. We reduce the arbitrariness associated with the choice of foliation prevailing in deterministictic clock models, and Equation \eqref{eq:guagecond} must be preserved under the dynamics generated by the total Hamiltonian.

As per the Dirac–Bergmann algorithm \cite{brown2022singular}, the temporal preservation of the gauge condition leads to a secondary constraint via the requirement
\begin{equation}
\dot{\chi}(x) = \{ \chi(x), H_{\text{tot}} \} \approx 0.
\end{equation}
Using the fundamental Poisson brackets and the explicit form of the matter Hamiltonian \(\mathcal{H}_\phi\), which is quadratic in \(\pi_{\phi}\), we also know that:
\begin{equation}
\dot{\chi}(x) = \int d^3y \, N(y) \{ \phi(x), \mathcal{H}(y) \}
= N(x) \frac{\delta \mathcal{H}(x)}{\delta \pi_\phi(x)},
\end{equation}
where one could assume the support of the delta function localizes the integral to $x = y$. 

The Hamiltonian constraint for a minimally coupled scalar field takes the form:
\begin{equation}
\mathcal{H}(x) = \mathcal{H}_{\text{grav}}(x) + \mathcal{H}_\phi(x),
\end{equation}
where
\begin{equation}
\mathcal{H}_\phi(x) = \frac{1}{2\sqrt{\gamma(x)}} \pi_\phi^2(x)
+ \frac{\sqrt{\gamma(x)}}{2} \gamma^{ij}(x) \partial_i \phi(x) \partial_j \phi(x)
+ \sqrt{\gamma(x)} V(\phi(x)).
\end{equation}The gravitational part \(\mathcal{H}_{\text{grav}}(x)\) will be invoked later in the section, as it is only required for analyzing the full constraint structure in ADM formalism. Since only the first term in \(\mathcal{H}_\phi\) depends on \(\pi_{\phi}\), taking the functional derivative with respect to $\pi_\phi(x)$ yields:
\begin{equation}
\frac{\delta \mathcal{H}(x)}{\delta \pi_\phi(x)} = \frac{\pi_\phi(x)}{\sqrt{\gamma(x)}}.
\end{equation}
Substituting into the consistency condition in Equation \eqref{eq:guagecond}, the preservation of the gauge implies:
\begin{equation}
\dot{\chi}(x) = N(x) \frac{\pi_\phi(x)}{\sqrt{\gamma(x)}} \approx 0.\label{eq:guagechoice}
\end{equation}

This condition yields two possible outcomes: either $N(x) = 0$, which corresponds to a degenerate foliation and halts any dynamical evolution, or $\pi_\phi(x) = 0$, which effectively freezes the scalar field and undermines its role as a relational clock. One might attempt to resolve this impasse by interpreting Equation \eqref{eq:guagecond} as a gauge-fixing equation that determines the lapse function $N(x)$ in terms of $\pi_\phi(x)$. However, such an approach is generally inadvisable in the presence of a non-vanishing field gradient $\partial_\mu \phi$, since the proper evolution must respect the slicing defined by the scalar field. In particular, the normal vector to constant-$\phi$ hypersurfaces is given by $n^\mu = \partial^\mu \phi / \sqrt{\partial^\nu \phi\, \partial_\nu \phi}$, and consistency with foliation geometry requires that Hamiltonian evolution proceed along this physical direction. Therefore, any determination of $N(x)$ must be compatible with this geometric foliation structure, ensuring that the scalar field evolves monotonically along $n^\mu$ and retains its function as an internal clock.

A more geometrically natural resolution arises by requiring that the scalar field evolves monotonically along the hypersurface normal. The relevant requirement is that the Lie derivative of $\phi$ along $n^\mu$ should satisfy \(\mathcal{L}_Z \phi = Z^\mu \partial_\mu \phi = N\, n^\mu \partial_\mu \phi = N \sqrt{\partial^\mu \phi \partial_\mu \phi} = 1,\) such that $\phi$ advances by one unit per unit coordinate time $\tau$. Solving for the lapse function, we obtain:
\begin{equation}
N(x) = \frac{1}{\sqrt{\partial^\mu \phi(x) \partial_\mu \phi(x)}}. \label{eq:Lpasechoice}
\end{equation}
By Substituting the expression for $\pi_\phi$ from Equation \eqref{eq:guagechoice}:
\begin{equation}
\pi_\phi(x) = \sqrt{\gamma(x)}\, \partial^0 \phi = \sqrt{\gamma(x)}\, \frac{1}{N(x)}\, \partial_t \phi,
\end{equation}
and combining the above expression with Equation \eqref{eq:Lpasechoice}, we obtain:
\begin{equation}
N(x) = \frac{\pi_\phi(x)}{\sqrt{\gamma(x)}\, \partial_\mu \phi(x) \partial^\mu \phi(x)}. \label{eq:lapse-from-clock}
\end{equation} under the assumption that \(\phi\) acts as a relational clock field in canonical gravity. A similar form where the lapse becomes dynamically determined by the momentum conjugate to a matter clock field has been derived in the Brown–Kuchař dust model \cite{brown1995dust, kucha1991gaussian}, and in loop‑cosmology deparametrizations \cite{husain2011dust, husain2012time}.

We further note that Equation \eqref{eq:lapse-from-clock} is a generalization following homogeneous cosmological models, where incoherent dust fields are used to define intrinsic spacetime coordinates and a true Hamiltonian emerges. It is the secondary constraint required to stabilize the gauge condition in Equation \eqref{eq:guagecond} under Hamiltonian evolution, such that it is generated along the physical direction defined by $\phi$. The resulting lapse function becomes a derived quantity rather than a freely specifiable Lagrange multiplier, tying the dynamics of the gravitational field directly to the evolution of the clock field.

However, for a minimally coupled scalar field \(\phi\), the configuration space consists of spatial 3-metrics \(\gamma_{ij}(x)\) and configurations of the scalar field \(\phi(x)\). The corresponding canonical momenta, \(\pi^{ij}(x)\) and \(\pi_\phi(x)\), are transformed into functional derivative operators. Following Dirac's prescription for constrained quantization, the Hamiltonian and momentum constraints are implemented as operator conditions on the state functional $\Psi[\gamma_{ij}, \phi]$:
\begin{equation}
\hat{\mathcal{H}}(x)\, \Psi[\gamma_{ij}, \phi] = 0, \qquad \hat{\mathcal{H}}_i(x)\, \Psi[\gamma_{ij}, \phi] = 0, \label{eq:WDform}
\end{equation}
which enforce invariance under spatial diffeomorphisms and deformations of the hypersurface. The quantized Hamiltonian constraint operator takes the form
\begin{equation}
\hat{\mathcal{H}}(x) = - \frac{16 \pi G \hbar^2}{\sqrt{\gamma(x)}}\, \mathcal{G}_{ijkl}(x)\, \frac{\delta^2}{\delta \gamma_{ij}(x) \delta \gamma_{kl}(x)}
- \frac{\sqrt{\gamma(x)}}{16 \pi G}\, {}^{(3)}\!R(x)
- \frac{\hbar^2}{2 \sqrt{\gamma(x)}}\, \frac{\delta^2}{\delta \phi(x)^2}
+ \sqrt{\gamma(x)}\, V(\phi(x)), \label{eq:WDqn}
\end{equation}
where $\mathcal{G}_{ijkl}(x)$ is the DeWitt supermetric,
\begin{equation}
\mathcal{G}_{ijkl}(x) = \frac{1}{2} \left( \gamma_{ik}(x)\gamma_{jl}(x) + \gamma_{il}(x)\gamma_{jk}(x) - \gamma_{ij}(x)\gamma_{kl}(x) \right),
\end{equation}
and $R$ (with superscript ${}^{(3)}$) is the scalar curvature of the induced 3-metric. Since the WD equation includes second functional derivatives with respect to $\phi$, it only resembles a Klein–Gordon-type equation once $\phi$ is identified as a relational clock. This reinterpretation emerges after gauge fixing the lapse function $N$ through the relational clock construction described earlier in the section. As the lapse functions as a Lagrange multiplier enforcing invariance under time reparametrizations, choosing a specific form of $N$ breaks this symmetry and effectively selects a preferred foliation. In doing so, one transitions from treating time as a coordinate artifact to identifying it with a dynamical field $\phi$, defined on the same manifold as the gravitational variables. The field $\phi$ then serves as a physical parameter labeling spacetime events, allowing us to reinterpret the Hamiltonian constraint not as a generator of gauge transformations but as a genuine generator of evolution. Operationally, this deparametrization transforms the WD equation into a Schrödinger- or Klein–Gordon-like evolution equation in the relational variable $\phi$ \cite{gielen2021hamiltonian, rovelli1990quantum}.

Having identified the scalar field $\phi$ as a relational time variable, we proceed to implement canonical quantization by promoting its conjugate momentum $\pi_\phi(x)$ to a functional derivative operator:
\begin{equation}
\pi_\phi(x) \longrightarrow -i \hbar\, \frac{\delta}{\delta \phi(x)}. \label{eq:kineticterm}
\end{equation}
Substituting Equation \eqref{eq:kineticterm} into the scalar field contribution to the Hamiltonian constraint (see Equation \eqref{eq:WDqn}) yields the kinetic term in the quantum theory:
\begin{equation}
\frac{\pi_\phi^2(x)}{2\sqrt{\gamma(x)}} \longrightarrow - \frac{\hbar^2}{2\sqrt{\gamma(x)}} \frac{\delta^2}{\delta \phi(x)^2}.
\end{equation}
Assuming that the scalar field $\phi(x)$ is spatially homogeneous—or, more generally, that it defines a global clock slicing such that $\phi(x) = \phi$ across the hypersurface $\Sigma$—we may suppress spatial dependence and replace the functional derivatives with ordinary derivatives:
\begin{equation}
\pi_\phi \longrightarrow -i \hbar\, \frac{d}{d \phi}, \quad \Psi[\gamma_{ij}, \phi(x)] \longrightarrow \Psi[\gamma_{ij}, \phi].
\end{equation}
Under this simplification, the WD equation reduces to a Schrödinger-type equation:
\begin{equation}
i \hbar\, \frac{d \Psi[\gamma_{ij}, \phi]}{d \phi} = \hat{\mathcal{H}}_{\mathrm{grav}}\, \Psi[\gamma_{ij}, \phi], \label{eq:schrodinger23}
\end{equation}
where the gravitational Hamiltonian operator takes the form:
\begin{equation}
\hat{\mathcal{H}}_{\mathrm{grav}} = - \frac{16 \pi G \hbar^2}{\sqrt{\gamma}}\, \mathcal{G}_{ijkl}\, \frac{\delta^2}{\delta \gamma_{ij} \delta \gamma_{kl}} - \frac{\sqrt{\gamma}}{16 \pi G}\, {}^{(3)}\!R + \sqrt{\gamma}\, V(\phi).
\end{equation}This procedure mirrors the standard deparametrization approach in minisuperspace quantum cosmology, wherein $\phi$ serves as an internal time parameter, and its conjugate momentum generates Schrödinger-type evolution for the gravitational wavefunction \cite{craig2010consistent}.

However, the scalar clock field is a relational construct—it derives its status as a time parameter only through the choice of gauge and reference frame. The emergence of Equation \eqref{eq:schrodinger23} is not a generic property of the WD equation, but rather a consequence of deparametrization within a reduced phase space. In the full theory, this picture is further complicated by quantum backreaction and higher-order gravitational corrections. When the scalar clock field exhibits intrinsic fluctuations, integrating out high-frequency (UV) gravitational degrees of freedom could lead to stochastic deformations—a phenomenon well established in the Einstein–Langevin framework. In the following section, we formalize this idea and derive a generalized WD equation in which stochasticity emerges as a physical trace of unresolved quantum gravitational fluctuations.

\subsection{Stochastic Clocks and Coarse-Grained Quantum Time}

In the previous sections, we discussed how the lapse function \(N(x)\) becomes dynamically tied to the momentum of the scalar clock field via Equation \eqref{eq:lapse-from-clock}. This relation enforces a foliation aligned with hypersurfaces of constant $\phi$, allowing for a deparametrized evolution in which $\phi$ serves as a relational time variable.  However, this construction rests on the assumption that \(\phi\) is a smooth, monotonic field with a well-defined conjugate momentum throughout the entire spatial hypersurface. While adequate at the classical or semiclassical level, such assumptions become untenable in a fully quantum gravitational regime, where the clock field itself interacts with—and is influenced by—the quantum geometry it probes.

A more complete picture emerges from the framework of stochastic gravity, which generalizes semiclassical gravity by incorporating quantum fluctuations of the matter stress-energy tensor as sources for metric perturbations. This program, extensively developed by Hu, Verdaguer, and collaborators (see \cite{hu1992quantum, hu1995fluctuation, hu2004induced} as well as contributions from Calzetta and Verdaguer), utilizes open quantum system techniques and nonequilibrium statistical field theory. Inspired by this approach, we extend the logic to the scalar clock field itself, treating it not as a classical background but as a quantum subsystem coupled to high-frequency (UV) gravitational degrees of freedom. Tracing out these short-wavelength metric modes generates stochastic noise in the clock dynamics via the influence functional formalism \cite{calzetta2009nonequilibrium}. The coarse-grained clock field thus acquires the structure of a stochastic process:
\begin{equation}
\phi(x) = t + t'(x), \label{eq:stochastc25}
\end{equation}
where \( t \) is the mean relational time and \( t'(x) \) represents coarse-grained, stochastic fluctuations.

To see this explicitly, consider the full wavefunctional \( \Psi[\gamma_{ij}, \phi, h_{ij}^{\mathrm{UV}}] \), where \( h_{ij}^{\mathrm{UV}} \) denotes high-frequency metric fluctuations. The total density matrix is then coarse-grained over these UV modes, yielding a reduced influence functional for the low-energy sector. The resulting effective action for the clock field becomes non-unitary and includes imaginary terms:
\begin{equation}
S_{\text{eff}}[\phi] = S_{\text{cl}}[\phi] + \int d^4x\, \xi(x)\, \phi(x), \qquad \langle \xi(x)\xi(y) \rangle = N(x, y),
\end{equation}
where \( \xi(x) \) is a Gaussian noise source and \( N(x, y) \) is the influence functional–derived noise kernel. This yields a Langevin-type effective field equation for the clock:
\begin{equation}
\square \phi(x) + V'(\phi) = \xi(x), \label{eq:clock-langevin}
\end{equation}
making the stochastic fluctuations in \( \phi \) a direct consequence of integrating out short-scale gravitational degrees of freedom.

In our model, the stochastic field is taken as a zero-mean Gaussian process with correlations:
\begin{equation}
\langle t'(x) \rangle = 0, \qquad \langle t'(x) t'(y) \rangle = G(x,y),
\end{equation}
where \( G(x,y) \sim N(x, y) \) induce foliation-level uncertainties. These kernels are often used in curved-space quantum field theory \cite{calzetta2001coarse} in accordance with the imaginary part of the influence action and have been extensively analyzed in contexts of cosmological decoherence \cite{roura2008cosmological}.

The stochastic fluctuation perturbs the normal vector to the constant-\( \phi \) hypersurfaces, and hence the lapse function in Equation~\eqref{eq:lapse-from-clock}. Writing:
\begin{equation}
\partial_\mu \phi(x) = \delta^0_\mu + \partial_\mu t'(x),
\end{equation}
and expanding the norm to leading order yields:
\begin{equation}
\sqrt{\partial_\mu \phi\, \partial^\mu \phi} \approx 1 + \frac{1}{2} \left( \partial_\mu t' \partial^\mu t' \right).
\end{equation}
So the lapse becomes:
\begin{equation}
N(x) \approx \frac{\pi_\phi(x)}{\sqrt{\gamma(x)} \left( 1 + \partial_\mu t' \partial^\mu t' \right)}.
\end{equation}
where an inverse dependence on $t'$ introduces stochastic corrections into the lapse, which propagate into the Hamiltonian constraint through the evolution vector $Z^\mu = N n^\mu + N^i e^\mu_i$. Subsequent fluctuation in the clock induces a corresponding noise in the lapse and therefore in the full Hamiltonian constraint.

At the operator level, this modifies the WD equation as:
\begin{equation}
\hat{\mathcal{H}}, \Psi \quad \longrightarrow \quad \left( \hat{\mathcal{H}} + \nabla_i \left( \frac{\xi^i(x)}{\sqrt{\gamma(x)}} \right) \right)\Psi = 0,
\end{equation}
where \( \xi^i(x) \) is a stochastic vector field, $\hat{\mathcal{H}}$ is the canonical Hamiltonian constraint operator and $\nabla_i$ denotes the spatial covariant derivative compatible with the three-metric $\gamma_{ij}$. The correlation function could be derived from the clock kernel:
\begin{equation}
\langle \xi^i(x) \rangle = 0, \qquad \langle \xi^i(x) \xi^j(y) \rangle = \tilde{G}^{ij}(x,y).
\end{equation}
These vector perturbations are analogous to “noise currents” stemming from short-scale matter fluctuations, and $\tilde{G}^{ij}(x,y)$ is a symmetric noise kernel analogous to those appearing in the Einstein–Langevin formalism of stochastic gravity. Physically, these terms modify the Hamiltonian constraint by introducing diffusion-like effects sourced by relational clock fluctuations, while preserving the constraint structure in expectation. 

The hypersurface-deformation algebra remains closed in expectation:
\begin{equation}
\left\langle \left[ \hat{\mathcal{H}}(x) + \nabla_i \left( \frac{\xi^i(x)}{\sqrt{\gamma(x)}} \right), \, \, \hat{\mathcal{H}}(y) + \nabla_j \left( \frac{\xi^j(y)}{\sqrt{\gamma(y)}} \right) \right] \right\rangle
= i \hbar, \hat{\mathcal{H}}^i(x), \partial_i \delta^3(x - y) + \cdots, \label{eq:basiseqn}
\end{equation} This expression does not represent a fundamental breakdown of constraint closure, but rather an effective deformation induced by coarse-graining. The emergence of relational time is not characterized as a classical slicing condition, but instead as a stochastic structure arising from the tracing over quantum gravitational degrees of freedom. The resulting WD equation is deformed by relational noise, in which coarse-grained relational observables evolve in a probabilistic, yet symmetry-consistent, fashion. 

\section{Stochastic Evolution and Diffusion in Clock Time}
In this section, we show that quantum fluctuations within a stochastic clock field cause a diffusion-like distortion of time evolution in canonical quantum gravity. These fluctuations result in a statistical broadening of the wavefunctional across neighboring clock slices, even in the absence of external noise. Starting from the WD equation, we derive a functional diffusion equation for the ensemble-averaged state. In a homogeneous cosmological setting, this reduces to a Langevin-type Schrödinger equation in minisuperspace, where relational time acquires an intrinsic quantum stochasticity.

\subsection{Diffusion Structure from Fluctuating Clocks}

We begin by considering how the stochastic fluctuation $t'(x)$ enters the wavefunctional. As the classical clock field is perturbed as shown in Equation \eqref{eq:stochastc25}, the state becomes a stochastic functional:
\begin{equation}
    \Psi[\gamma_{ij}, \phi] = \Psi[\gamma_{ij}, t + t'(x)],
\end{equation} were we expand it as a functional Taylor series around the mean clock time $t$:
\begin{equation}
\Psi[\gamma_{ij}, \phi] = \Psi[\gamma_{ij}, t + t'(x)] 
= \Psi[\gamma_{ij}, t] 
+ \int d^3x\, t'(x)\, \frac{\delta \Psi}{\delta \phi(x)} 
+ \frac{1}{2} \int d^3x\, d^3y\, t'(x) t'(y) \frac{\delta^2 \Psi}{\delta \phi(x) \delta \phi(y)} + \cdots.
\end{equation}In the above expression, the linear term indicates the first-order sensitivity of the quantum state to local foliation shifts, while the quadratic term shows the cumulative effect of pairwise correlations between fluctuations at different spatial points. The which will dominate the stochastic average due to the assumed Gaussianity.

Taking the statistical ensemble average $\langle \cdot \rangle$ over realizations of the stochastic process, and noting that $\langle t'(x) \rangle = 0$, we find:
\begin{equation}
\langle \Psi[\gamma_{ij}, \phi] \rangle 
= \Psi[\gamma_{ij}, t] 
+ \frac{1}{2} \int d^3x\, d^3y\, G(x, y) \frac{\delta^2 \Psi}{\delta \phi(x) \delta \phi(y)} + \cdots.
\end{equation}
Here, the second-order functional derivative term, proportional to $G(x,y)$, acts as a diffusion operator over the clock field $\phi(x)$. This implies that even if the fundamental theory is deterministic at the level of the WD equation, the ensemble-averaged, coarse-grained quantum state $\langle \Psi \rangle$ evolves according to a non-unitary, diffusive effective theory. For instance, consider a classical analogy: the microscopic dynamics of a gas are governed by deterministic Newtonian trajectories of individual molecules. However, when averaged over microscopic degrees of freedom, the emergent macroscopic behavior is described by diffusion equations (e.g., heat or particle diffusion). Similarly, in the quantum gravitational context, averaging over spatial clock fluctuations generates a functional diffusion structure in relational time \cite{ellis2000quantum}.

The ensemble-averaged state thus satisfies an effective quantum constraint:
\begin{equation}
\left( \hat{\mathcal{H}}_{\text{grav}} - \frac{\hbar^2}{2} \int d^3x\, d^3y\, G(x, y)\, \frac{\delta^2}{\delta \phi(x) \delta \phi(y)} \right) \langle \Psi \rangle = 0,
\label{eq:diffusive-WDW}
\end{equation}
which generalizes the WD constraint into a nonlocal functional diffusion equation in the clock field. The second-order functional derivatives with respect to \(\phi(x)\) capture the stochastic spread of the quantum state over neighboring relational time slices. A similar functional diffusion form has been discussed in the context of open quantum systems in \cite{breuer2002theory}, where noise arising from coupling to unobserved degrees of freedom induces non-unitary evolution. However, it is important to mention that when we consider \(\phi(x)\) itself a quantum field subject to environmental entanglement and decoherence, the foliation it specifies becomes imprecise. This imprecision could manifest as a (nonlocal) functional derivative term in the model, structurally different from the diffusion term in a Fokker-Planck equation in emergent stochastic gravity models.

The diffusion-like deformation of the quantum constraint could arise from coarse-graining over UV degrees of freedom \cite{ronco2016uv, liu2024path}. Formally, this can be justified through influence functional techniques in the spirit of Feynman and Vernon \cite{feynman1963theory, calzetta1988nonequilibrium}, adapted to canonical quantum gravity. In such approaches, the full theory is partitioned into resolved (IR) and unresolved (UV) modes:
\begin{equation}
\Psi[\gamma_{ij}^{\text{IR}}, \phi] = \int \mathcal{D} \gamma_{ij}^{\text{UV}}\, e^{i S[\gamma_{ij}^{\text{IR}}, \gamma_{ij}^{\text{UV}}, \phi]/\hbar},
\end{equation}
and tracing over the UV sector induces both decoherence and noise in the reduced theory. For instance, in the work by \cite{agon2018coarse}, this dynamics is illustrated using a model system characterized by a simple hierarchy of energy gaps, specifically \(\Delta E_{UV} > \Delta E_{IR}\). In this context, the coupling between high-energy and low-energy degrees of freedom is treated using second-order perturbation theory. Consequently, the kernel can be interpreted as the noise kernel associated with the stochastic response of the unresolved sector to the clock-field slicing. In general, its structure is determined by the spectral density of these unresolved fluctuations. More refined models may relate $G(x,y)$ to the Hadamard or Schwinger–Keldysh noise kernels arising in semiclassical gravity and stochastic inflation \cite{burgess2016open}. This will be addressed in future work and is not included in the preliminary development of the model presented here.

\subsection{Minisuperspace Limit and Langevin Schrödinger Evolution}

To extract physically interpretable predictions from the functional diffusion structure derived in the previous subsection, we now specialize to a homogeneous cosmological setting. This truncation leads to a minisuperspace model, in which both the spatial 3-metric $\gamma_{ij}(x)$ and the scalar clock field $\phi(x)$ are assumed to be spatially uniform across each hypersurface $\Sigma_t$. That is, we set:
\begin{equation}
\gamma_{ij}(x) \equiv a^2(t)\, \bar{\gamma}_{ij}, \qquad \phi(x) \equiv \phi(t),
\end{equation}
where $a(t)$ is the scale factor and $\bar{\gamma}_{ij}$ is a fixed fiducial metric on a compact spatial manifold (such as a 3-sphere or a flat 3-space). The degrees of freedom are now finite-dimensional: the configuration space reduces to $\mathcal{C} = \{a, \phi\}$, and the wavefunctional simplifies to a wavefunction:
\begin{equation}
\Psi[\gamma_{ij}, \phi] \rightarrow \psi(a, \phi).
\end{equation}

Within this reduced setting, the WD equation becomes an ordinary functional differential equation in a finite-dimensional minisuperspace. The stochastic fluctuations in the slicing variable $t'(x)$ representing clock-field inhomogeneities, reduce to a single global fluctuation $t'$, modeled as a random variable:
\begin{equation}
\langle t' \rangle = 0, \qquad \langle {t'}^2 \rangle = \sigma^2,
\end{equation} where \(\sigma^2\) is the integrated variance of spatial clock fluctuations, capturing the effective noise amplitude under the minisuperspace approximation \cite{schander2021backreaction}. Expanding the wavefunction in powers of $t'$, as in the functional case, we obtain:
\begin{equation}
\psi(\phi) = \psi(t + t') = \psi(t) + t' \frac{d\psi}{dt} + \frac{1}{2} {t'}^2 \frac{d^2 \psi}{dt^2} + \cdots.
\end{equation}
Since \(t'\) models stochastic fluctuations of the relational clock field around its classical mean, we compute the ensemble average over its realizations to extract the effective, coarse-grained evolution of the quantum state. Taking the ensemble average over the stochastic variable $t'$ yields:
\begin{equation}
\langle \psi(\phi) \rangle = \psi(t) + \frac{\sigma^2}{2} \frac{d^2 \psi}{dt^2} + \cdots.
\end{equation}This shows that stochastic fluctuations in the clock variable lead to a broadening of the quantum state in relational time, generating higher-order derivative corrections to Schrödinger evolution. At leading order, the averaged wavefunction satisfies:
\begin{equation}
i \hbar \frac{d \langle \psi \rangle}{dt} = \mathcal{H}_{\text{eff}} \langle \psi \rangle - \frac{i \hbar \sigma^2}{2} \frac{d^3 \langle \psi \rangle}{dt^3} + \cdots,
\end{equation}
where $\mathcal{H}_{\text{eff}} = -\hbar^2 \partial_a^2 + U(a, \phi)$ is the minisuperspace gravitational Hamiltonian and $U(a, \phi)$ is the effective potential from spatial curvature and scalar field terms \cite{isichei2023minisuperspace, venturi1990minisuperspace}. 

While Langevin-type corrections have appeared in various formulations of stochastic quantum mechanics \cite{gevorkyan2004exactly, okamoto1990stochastic, olavo2012foundations, roncadelli1991langevin, bargueno2014generalized}, and Fokker–Planck dynamics have been explored in gravitational settings \cite{erlich2018stochastic, erlich2022first}, our approach is distinct in its intrinsic origin of stochasticity. In the present framework, individual realizations of the quantum state obey a Langevin-deformed Schrödinger equation:
\begin{equation}
i \hbar \frac{d \psi}{d\phi} = \mathcal{H}_{\text{eff}}\, \psi + \hbar\, \eta(\phi),
\label{eq:langevin-schro}
\end{equation}
where $\eta(\phi)$ is a real-valued stochastic process with:
\begin{equation}
    \langle \eta(\phi) \rangle = 0, \qquad \langle \eta(\phi)\, \eta(\phi') \rangle = G(\phi, \phi').
\end{equation}Here, the stochastic term $\eta(\phi)$ acts analogously to a fluctuating force in Langevin dynamics. In the limit $G(\phi, \phi') \to 0$, relational time evolution reverts to standard unitary Schrödinger evolution--a regime in which $\phi$ behaves as a sharply defined global clock. Else, it quantify the strength of intrinsic fluctuations sourced by high-frequency gravitational modes integrated out in the coarse-graining procedure; a detailed spectral form depends on the cutoff and graviton mode structure.

In this picture, the quantum evolution is no longer governed by a unitary Schrödinger equation but rather by a non-unitary ensemble of stochastic trajectories. For the latter case, the average evolution can be described by a Fokker-Planck-type master equation. Specifically, if $\rho(\phi) := \langle \psi(\phi)\psi^*(\phi) \rangle$ denotes the ensemble-averaged density matrix, then its evolution follows:
\begin{equation}
\frac{d \rho}{d\phi} = -\frac{i}{\hbar} [\mathcal{H}_{\text{eff}}, \rho] + \frac{1}{\hbar^2} \int d\phi'\, G(\phi, \phi')\, \delta(\phi - \phi')\, \left( \psi\, \psi^* \right)'' + \cdots, \label{eq:secondderi}
\end{equation} where $(\psi \psi^*)'' := {d^2}/{d\phi^2} (\psi(\phi) \psi^*(\phi))$ denotes the second derivative of the pure-state density matrix with respect to relational time. This term captures the rate at which stochastic spreading affects the quantum amplitude distribution across neighboring clock slices. 

For a local diffusion structure, one may assume that the two-point correlator $G(\phi, \phi')$ is sharply peaked around $\phi = \phi'$—i.e., that the noise is delta-correlated in relational time, analogous to the white-noise limit in stochastic differential equations \cite{evans2012introduction}. The nonlocal noise term can be approximated by:
\begin{equation}
    \frac{d \rho}{d\phi} = -\frac{i}{\hbar} [\mathcal{H}_{\text{eff}}, \, \rho] + \frac{1}{2} \frac{d^2}{d\phi^2} \left[ D(\phi), \, \rho \right] + \cdots,
\end{equation}where $D(\phi) := \hbar^{-2} G(\phi, \phi)$ is an effective diffusion coefficient. Symmetrizing the second derivative operator ensures that the diffusion term maintains time-reversal invariance at the microscopic level \cite{hasegawa1980stochastic}. This corresponds to the Stratonovich interpretation of stochastic integrals, which maintains the usual chain rule under noise averaging \cite{stratonovich1966new, moon2014interpretation}, where ambiguities in operator ordering are resolved by treating the stochastic correction as a symmetric adjustment to the deterministic generator.

Moreover, the intrinsic stochastic dynamics discussed here is conceptually distinct from the approach developed by Hu and Verdaguer, where non-unitary evolution arises from coarse-graining over quantum matter fields or trans-Planckian modes. In their framework, stochasticity is extrinsic, modeled as noise sourced by the environment within a system–environment paradigm of semiclassical gravity (see \cite{campos1994semiclassical} and refs therein). However, the present framework realizes stochasticity as a feature of relational quantization. Thus, one may regard \( G(\phi, \phi') \) as fundamentally of quantum gravitational origin, encoding the influence functional associated with the underlying stochastic geometry. Otherwise, canonical quantization necessitates a fundamentally probabilistic temporal structure, wherein time acquires an irreducible quantum uncertainty, manifesting as a diffusion process in the evolution of the wavefunction.

\section{Physical Origin of Stochastic Clocks from Graviton Coarse-Graining}

Building on the functional diffusion structure derived in the previous section, we now investigate the microscopic origin of stochasticity in relational time. While the stochastic deformation of quantum evolution was introduced via slicing fluctuations of the clock field, such fluctuations must themselves arise from physical processes within quantum gravity \cite{calzetta1994noise}. In this section, we show that these fluctuations could emerge from coarse-graining the UV sector of gravity. This provides a first-principles origin for the stochastic clock dynamics introduced earlier, rooted in the quantum structure of spacetime itself.

\subsection{Clock Current Conservation and Foliation Stability}

A consistent interpretation of \(\phi(x)\) as a relational clock in canonical quantum gravity indicates that its evolution must create a stable and coherent foliation of spacetime \cite{Isham1993}.  That is, the level sets \(\Sigma_\phi\) defined by \(\phi(x) = \text{const.}\) must form a well-behaved slicing of spacetime that respects the causal and geometrical structure of the background. In particular, the scalar’s momentum current $\pi^\mu := \nabla^\mu \phi$ must form a timelike, divergence-free vector field that respects the causal structure of the background geometry. 

The Lagrangian density for a minimally coupled massless scalar field in a curved background is given by:
\begin{equation}
\mathcal{L}_\phi = \frac{1}{2} g^{\mu\nu} \nabla_\mu \phi\, \nabla_\nu \phi = \frac{1}{2} (\partial \phi)^2.
\end{equation}
yielding the wave equation:
\begin{equation}
\Box \phi = \nabla^\mu \nabla_\mu \phi = 0.
\end{equation}
The associated energy-momentum tensor is:
\begin{equation}
T_{\mu\nu} = \nabla_\mu \phi \nabla_\nu \phi - \frac{1}{2} g_{\mu\nu} (\nabla^\lambda \phi \nabla_\lambda \phi),
\end{equation}
and the canonical momentum current \(\pi^\mu := \nabla^\mu \phi\) satisfies the conservation law:
\begin{equation}
\nabla_\mu \nabla^\mu \phi = \Box \phi = 0,
\end{equation}
when the scalar field satisfies its classical equations of motion.

This conserved, timelike flow provides a natural structure for constructing a globally stable slicing of spacetime. Within the ADM formalism, the normal vector to the hypersurface \(\Sigma_\phi\) is given by:
\begin{equation}
n^\mu = \frac{\pi^\mu}{\sqrt{\pi^\nu \pi_\nu}},
\end{equation}
and the divergence-free condition ensures that these slices evolve consistently across space and time. In the latter case, relational time corresponds to the parameterization along integral curves of $\pi^\mu$, making the scalar field an operational clock that defines physical time via its own dynamics.

However, this idealized construction is extremely sensitive to perturbations. Even a small mass term in the Lagrangian:
\begin{equation}
\mathcal{L}_\phi = \frac{1}{2} (\partial \phi)^2 - \frac{1}{2} m_0^2 \phi^2,
\end{equation}
modifies the dynamics to the Klein–Gordon equation:
\begin{equation}
\Box \phi - m_0^2 \phi = 0,
\end{equation}
and induces a source term in the momentum current:
\begin{equation}
\nabla_\mu \pi^\mu = - m_0^2 \phi.
\end{equation}
This loss of current conservation implies that the scalar field no longer defines a consistent foliation. Physically, this leads to a breakdown of simultaneity across space: the rate at which "clock time" progresses becomes spacetime-dependent, introducing temporal decoherence and disrupting the relational slicing. Figure \ref{fig:divfree} illustrates this using a Gaussian-modulated oscillatory field to show how the presence of a mass scale $m_0$ acts as a source term in the effective fluid description of the clock field, consequently breaking the conservation of the clock current. In this context, we can interpret the fluctuations in the foliation of stochastic clocks as the distinctive physical imprint of the conserved flow of \(\phi\), rather than viewing it as an arbitrary sea of metric foam or fuzziness.
 
\begin{figure*}[ht]
    \centering
    \includegraphics[width=\linewidth]{streamplot_conservation_comparison.png}
    \caption{\small 2D streamline plot of the clock momentum density current \(\pi^\mu(x)\), illustrating the divergence \(\nabla_\mu \pi^\mu\) in a spatial slice. Left: For \(m_0 = 0\), \(\nabla_\mu \pi^\mu \approx 0\) indicates conservation of the current, with flow lines forming closed, divergence-free configurations. Right: For \(m_0 \neq 0\), \(\nabla_\mu \pi^\mu \neq 0\) signifies broken conservation. Warm colors (orange-red) represent regions of positive divergence (source-like behavior), while cool colors (blue) indicate negative divergence (sink-like behavior).}
    \label{fig:divfree}
\end{figure*}

For the massless case, the current $\pi^\mu$ defines a globally divergence-free, irrotational flow, ideal for generating stable slices. In contrast, the massive field exhibits localized sources and sinks in its flow, undermining the integrity of the slicing. Importantly, even in the absence of an explicit mass term, similar violations can arise from the quantum gravitational sector. Specifically, fluctuations in the underlying metric—particularly unresolved graviton modes—can induce effective sources in the current conservation equation. These effective violations appear as stochastic deformations of the constraint algebra depending on the energy density \cite{bojowald2008loop, mielczarek2014loop, amelino2017spacetime, ronco2016uv}, and modifications in dispersion relations.

In contrast to deterministic foliations, the clean slicing provided by the classical scalar field becomes "fuzzy" under quantum corrections, and the evolution of $\phi$ acquires a stochastic component—an observable subject to quantum diffusion due to gravitational coarse-graining \cite{bodendorfer2017state}. In the next section, we derive this result explicitly using the path integral formalism, where graviton backreaction appears as an effective Langevin-type force deforming the constraint algebra.

\subsection{Graviton-Induced Noise from Coarse-Graining}

In the previous section, we argued that the stability of a relational scalar clock requires the conservation of its momentum current. However, quantum gravitational backreaction can induce effective violations of this conservation law. In this section, we show that such stochasticity arises naturally by integrating out the physical graviton modes in the path integral formulation, leading to the deformation of the clock dynamics consistent with the canonical constraint structure in Equation \eqref{eq:basiseqn}.

We begin with the Einstein–Klein–Gordon (EKG) system on a four-dimensional Lorentzian manifold \(M\), described by the classical action:
\begin{equation}
S[g_{\mu\nu}, \phi] = \int_M d^4x\, \sqrt{-g} \left( \frac{1}{16\pi G} R - \frac{1}{2} g^{\mu\nu} \nabla_\mu \phi \nabla_\nu \phi \right),
\end{equation}
where \(R\) is the Ricci scalar of the spacetime metric \(g_{\mu\nu}\) and \(\phi\) is a minimally coupled massless scalar field. The corresponding quantum theory is described by the generating functional:
\begin{equation}
\mathcal{Z} = \int \mathcal{D}g_{\mu\nu}\, \mathcal{D}\phi\; \Delta_{\text{FP}}[N, N^i] \, e^{i S[g_{\mu\nu}, \phi]/\hbar},
\label{eq:Zfull}
\end{equation}
where \(\Delta_{\text{FP}}[N, N^i]\) is the Faddeev–Popov determinant associated with gauge-fixing the lapse function \(N\) and shift vector \(N^i\) in the ADM decomposition. This procedure removes redundant contributions from coordinate diffeomorphisms and enables us to isolate the true dynamical degrees of freedom—namely, the transverse-traceless (TT) graviton modes and the scalar field—by eliminating gauge artifacts from the functional integral.

To isolate the quantum gravitational fluctuations relevant for backreaction, we decompose the spatial 3-metric into a background and a perturbation:
\begin{equation}
\gamma_{ij}(x) = \bar{\gamma}_{ij}(x) + \delta\gamma_{ij}(x),
\end{equation}
where \(\bar{\gamma}_{ij}\) denotes a classical background geometry satisfying the semiclassical Einstein equations, and \(\delta \gamma_{ij}\) represents the quantum fluctuations. Under the ADM gauge-fixing implemented via $\Delta_{\text{FP}}[N,N^i]$, the only remaining dynamical components are the TT perturbations $\delta \gamma^{\text{TT}}_{ij}$, corresponding to the two propagating physical polarizations of the graviton \cite{dewitt1967quantum, halliwell1991wave}. Scalar and longitudinal metric modes are either constrained by the Hamiltonian and momentum constraints or canceled by ghost contributions at one-loop, and hence decouple from the scalar field sector at leading order in perturbation theory \cite{esposito2012euclidean}.

With the metric decomposition in hand, we now expand the total action around the classical background geometry. To leading order, this yields:
\begin{equation}
S[g_{\mu\nu}, \phi] = S[\bar{g}_{\mu\nu}, \phi] + \delta S[\delta \gamma_{ij}, \phi] + \mathcal{O}(\delta \gamma^3),
\end{equation}
where \(\delta S\) includes all terms quadratic in the TT graviton perturbations and their couplings to the scalar field. Linear terms vanish under the assumption that $\bar{g}_{\mu\nu}$ satisfies the background equations of motion. To derive the effective dynamics of $\phi$, we integrate out the TT graviton modes at one-loop:
\begin{equation}
e^{i S_{\text{eff}}[\phi]/\hbar} = \int \hat{\mathcal{D}} \delta \gamma_{ij}^{\text{TT}} \, e^{i S[\bar{g} + \delta \gamma, \phi]/\hbar}.
\end{equation}
The result is a nonlocal influence functional, encoding the leading-order backreaction of quantum gravitational fluctuations on the scalar clock field:
\begin{equation}
S_{\text{eff}}[\phi] = S[\bar{g}, \phi] + \frac{i\hbar}{2} \int d^4x\, d^4y\, O^{ij}[\phi(x)]\, G_{ijkl}(x,y)\, O^{kl}[\phi(y)] + \cdots.
\label{eq:Seff}
\end{equation}
Here, $O^{ij}[\phi(x)] $ represents the metric response of the scalar Hamiltonian density, acting as the gravitational ‘backreaction channel’, while $G_{ijkl}(x,y)$ is the graviton two-point function (propagator) in the chosen gauge (e.g., de Donder or harmonic). This effective action can be viewed as a semiclassical realization of the Feynman–Vernon influence functional in a gravitational context, capturing decoherence and stochasticity induced by unresolved geometric degrees of freedom.

The influence term in Equation~\eqref{eq:Seff} is bilocal in the scalar field configuration and arises from contracting the graviton propagator $G_{ijkl}(x, y)$ with functional metric derivatives of the scalar Hamiltonian. While this captures the non-Markovian backreaction of TT graviton modes, it is analytically intractable due to the bilocal convolution. To reformulate this interaction into a stochastic local coupling, we perform a Hubbard–Stratonovich transformation, rewriting the nonlocal quadratic influence term as a Gaussian integral over an auxiliary tensor field \(\xi^{ij}(x)\):
\begin{equation}
e^{\frac{i\hbar}{2} \int O G O} = \int \mathcal{D} \xi^{ij}\, \exp \left[ i \int d^4x\, \xi^{ij}(x)\, O_{ij}[\phi(x)] - \frac{1}{2\hbar} \int d^4x\, d^4y\, \xi^{ij}(x)\, K^{-1}_{ijkl}(x,y)\, \xi^{kl}(y) \right],
\end{equation}
where \(K^{-1}_{ijkl}(x,y)\) is the inverse of the graviton propagator kernel \(G_{ijkl}(x,y)\), assumed to be positive definite (or suitably regulated). This transformation trades the bilocal nonlinearity for a linear but stochastic coupling to an auxiliary tensor field. The auxiliary field $\xi^{ij}(x)$ now appears as a Gaussian-distributed stochastic source with covariance:
\begin{equation}
    \langle \xi^{ij}(x)\, \xi^{kl}(y) \rangle = \hbar\, G^{ij\,kl}(x, y),
\end{equation}determined entirely by the TT graviton correlator. Through its coupling to $O_{ij}[\phi(x)]$, this field mediates the effective backreaction of the unresolved gravitational modes on the scalar sector. In this way, $\xi^{ij}(x)$ realizes a covariant and gauge-consistent spacetime noise term, and the scalar field dynamics inherit a Langevin-type structure driven by fluctuations in the gravitational vacuum.

To make the structure of the coupling more explicit, we now compute the operator $O^{ij}[\phi]$ in terms of canonical variables within the ADM decomposition. In this formalism, the scalar field contributes to the Hamiltonian constraint through both its conjugate momentum $\pi_\phi$ and spatial derivatives. The scalar Hamiltonian density takes the form:
\begin{equation}
\mathcal{H}_\phi = \frac{\pi_\phi^2}{2\sqrt{\gamma}} + \sqrt{\gamma} \, \gamma^{ij} \partial_i \phi \partial_j \phi.
\end{equation}
where $\gamma_{ij}$ is the spatial 3-metric on the hypersurface $\Sigma_t$. The operator $O^{ij}[\phi]$, defined as the functional derivative of $\mathcal{H}_\phi$ with respect to $\gamma_{ij}$, evaluates to:
\begin{equation}
O^{ij}[\phi] = \frac{\delta \mathcal{H}_\phi}{\delta \gamma_{ij}} = -\frac{1}{2} \left( \frac{\pi_\phi^2}{\sqrt{\gamma}} \, \gamma^{ij} + \sqrt{\gamma} \left[ \gamma^{ik} \gamma^{jl} - \frac{1}{2} \gamma^{ij} \gamma^{kl} \right] \partial_k \phi \partial_l \phi \right) + \cdots,
\end{equation}
where the ellipsis denotes subleading terms, including total derivatives and higher-order corrections in the metric perturbations. This expression shows that the stochastic source $\xi^{ij}(x)$ couples nontrivially to both the kinetic and gradient energy densities of the scalar field, such that the resulting dynamics is manifestly nonlocal and gauge-invariant at leading order.

To isolate the influence of slicing fluctuations and simplify the effective coupling, we contract the stochastic tensor \(\xi^{ij}(x)\) over the metric and take its covariant divergence to define an effective vector field:
\begin{equation}
\xi^i(x) := \nabla_j \left( \gamma^{ij} \xi_{kl}(x) \gamma^{kl} \right).
\end{equation}
This contraction projects the graviton-induced noise into a vector density whose divergence yields a scalar of appropriate weight to enter the Hamiltonian constraint. Applying the Leibniz rule to the covariant derivative, the corresponding stochastic deformation of the constraint becomes:
\begin{equation}
\mathcal{H}_{\text{stoch}}(x) = \frac{1}{\sqrt{\gamma}} \nabla_i \xi^i(x) - \frac{1}{2} \xi^i(x) \nabla_i \log \gamma(x), \label{eq:conalge}
\end{equation}
where the resulting quantum constraint thus acquires a Langevin-type structure of the form:
\begin{equation}
    \left( \hat{\mathcal{H}}_{\text{grav}} + \hat{\mathcal{H}}_{\phi} + \hat{\mathcal{H}}_{\text{stoch}} \right)(x)\, \Psi[\gamma_{ij}, \phi] = 0.
\end{equation}Importantly, this structure preserves the closure of the constraint algebra in Equation \eqref{eq:basiseqn}. Each realization corresponds to a different coarse-grained configuration of the gravitational field, indicating the inherently probabilistic structure of relational time in quantum gravity.

Note that the emergence of a Langevin-type structure in the quantum Hamiltonian constraint represents more than just a technical modification. Rather than enforcing a single deterministic trajectory for the relational clock field, this stochastic constraint recognizes the influence of unresolved gravitational degrees of freedom, appearing as noise in \( \phi \)-based evolutions. This reinterprets time not as a fixed parameter but as an ensemble of stochastic histories shaped by quantum geometry.

\begin{figure}
\centering
\resizebox{0.8\columnwidth}{!}{%
\begin{tikzpicture}[
    node distance=1.2cm and 2.5cm,
    box/.style={draw, rounded corners, minimum width=3.5cm, minimum height=1.2cm, align=center, font=\small, fill=blue!10},
    midbox/.style={draw, rounded corners, minimum width=3.5cm, minimum height=1.2cm, align=center, font=\small, fill=orange!10},
    outbox/.style={draw, rounded corners, minimum width=4.1cm, minimum height=1.2cm, align=center, font=\small, fill=green!10},
    futurebox/.style={draw, rounded corners, minimum width=4.2cm, minimum height=1.2cm, align=center, font=\small, fill=gray!15},
    arrow/.style={-{Latex[length=2mm]}, thick}
]

\node[box] (clock) {Relational Scalar Clock\\$\phi(x)$};
\node[box, right=of clock] (ADM) {Canonical Constraints\\(ADM formalism)};
\node[box, right=of ADM] (WDW) {Wheeler--DeWitt Equation\\$\hat{\mathcal{H}} \Psi = 0$};

\node[midbox, below=of clock] (graviton) {Graviton Backreaction};
\node[midbox, below=of ADM] (coarse) {Coarse-graining\\(Influence functional)};
\node[midbox, below=of WDW] (stochclock) {Stochastic Clock Field};

\node[midbox, below=of stochclock] (deform) {Deformed Constraint\\$\hat{\mathcal{H}} + \mathcal{H}_{\text{stoch}}$};

\node[outbox, below=of deform] (evolution) {\\ \\ Ensemble-Averaged Evolution \\  $\langle \Psi \rangle$ };

\node[outbox, left=of deform] (langevin) {Langevin Noise Source\\$\xi^i(x)$};

\node[futurebox, below left=0.5cm and -0.5cm of langevin] (mini) {Minisuperspace FLRW \\ with $\xi$-deformation};

\node[futurebox, below=of langevin] (singularity) {Singularity Avoidance\\via Stochastic Suppression};

\node[futurebox, below right=0.5cm and -0.5cm of langevin] (decoh) {Anisotropic Models};

\draw[arrow] (clock) -- (ADM);
\draw[arrow] (ADM) -- (WDW);

\draw[arrow] (clock) -- (graviton);
\draw[arrow] (ADM) -- (coarse);
\draw[arrow] (WDW) -- (stochclock);

\draw[arrow] (graviton) -- (coarse);
\draw[arrow] (coarse) -- (stochclock);
\draw[arrow] (stochclock) -- (deform);
\draw[arrow] (deform) -- (evolution);

\draw[arrow] (deform.west) -- (langevin.east);
\draw[arrow] (langevin) -- (mini);
\draw[arrow] (langevin) -- (singularity);
\draw[arrow] (langevin) -- (decoh);

\end{tikzpicture}
}
\label{fig:stochflow}
\caption{Conceptual flow of the stochastic clock framework. Blue boxes denote well-established structures in the literature, orange boxes correspond to technical developments from stochastic gravity and open quantum systems, green boxes indicate our results, and gray boxes outline future extensions.}

\end{figure}

\section{Discussion}
This paper offers a sharp departure from traditional relational time models—including the Page–Wootters mechanism, unimodular time, and York time—which posit deterministic internal clocks but do not account for gravitational backreaction. While semiclassical influence functionals have previously been applied to metric perturbations, such formulations have not been coupled to the Hamiltonian constraint algebra or interpreted as defining relational clock dynamics. Our model thus extends the scope of canonical quantum gravity by operationalizing time as a statistical construct, with fluctuations sourced by geometric modes. This reframing provides a concrete mechanism for temporal diffusion at the quantum gravitational level, and resonates well with timeless formulations of quantum gravity \cite{barbour1994timelessnessfirst, barbour1994timelessness}, in which classical time arises only through correlations among physical degrees of freedom. Crucially, the emergence of stochasticity in our framework is not merely a phenomenological addition but a structural necessity: it maintains the closure of the modified constraint algebra and preserves foliation invariance under coarse-graining. Without such stochastic corrections, the relational clock would fail to consistently propagate quantum states across hypersurfaces in a diffeomorphism-invariant theory.

A stochastic clock, in this context, is a relational time variable—typically a scalar field—whose evolution is subject to intrinsic fluctuations arising from unresolved gravitational degrees of freedom. In contrast to classical or semiclassical treatments where the clock evolves deterministically, the stochastic clock acquires Langevin-type noise in its momentum, modeled here as a consequence of coarse-graining TT graviton modes. This leads to a deformation of the WD equation into a Fokker–Planck-type evolution, where the clock variable drives a diffusive—not unitary—evolution of the quantum state. As a result, hypersurfaces of constant \(\phi\)—representing equal clock time slices are no longer deterministically connected due to noise-induced foliation fluctuations. As emphasized in \cite{nelson1973construction, nelson2014stochastic}, such non-pathwise-connected clock hypersurfaces are not pathologies but manifestations of the intrinsic nonlocality of quantum gravitational time. Our formalism thus situates itself as a minisuperspace realization of this idea, where the ensemble average reproduces the usual relational evolution, while its two-point variance grows linearly in clock time.  Further research is needed to assess whether these stochastic corrections remain consistent in higher-dimensional or nonperturbative formulations of quantum gravity.

\section{Outlook}

This work conjectures that quantum fluctuations in the relational clock field deform the Hamiltonian constraint into a stochastic ensemble of quantum constraints. Within this framework, time ceases to be a sharply defined parameter and instead emerges probabilistically through foliation-dependent dynamics. A natural next step is to evaluate the full minisuperspace reduction of the stochastic constraint algebra in FLRW spacetimes. Preliminary indications suggest that the induced Langevin dynamics suppress the probability of small-scale configurations, offering a potential resolution mechanism for classical singularities. Quantitative predictions for the noise-averaged evolution of the scale factor, $\langle a(\phi) \rangle$, and its diffusion profile will be presented in future work, including a detailed comparison with the Erlich composite gravity framework. Beyond minisuperspace, open questions remain regarding the consistency of relational observables across distinct stochastic foliation histories. Whether diffusion in the clock variable reflects a fundamental quantum feature of spacetime or a coarse-graining artifact remains to be understood. Future investigations could also extend the formalism to anisotropic or inhomogeneous cosmologies—such as Bianchi I or Lemaître–Tolman–Bondi models—where directional degrees of freedom may interact nontrivially with the stochastic clock. These generalizations may offer new insights into how clock fluctuations manifest in realistic gravitational settings.

\section{Declaration of competing interest}
\begin{enumerate}
    \item Funding: This research received no external funding.
    \item Data Availability Statement: Not applicable.
    \item Conflicts of Interest: The author declares no conflict of interest.
\end{enumerate}

\printbibliography

\end{document}